\documentclass[journal]{IEEEtran}
\usepackage{framed,multirow}
\usepackage{url}
\usepackage{xcolor}
\usepackage{hyperref}
\usepackage{amssymb}
\usepackage{latexsym}
\usepackage{url}
\usepackage{xcolor}
\usepackage{hyperref}
\usepackage{subfigure} 
\usepackage{algorithm}
\usepackage{algorithmic}
\usepackage{setspace}
\usepackage{amsmath}
\usepackage{diagbox}
\usepackage{array}
\usepackage{pbox}
\usepackage{graphicx}
\newcolumntype{L}{>{\centering\arraybackslash}m{6cm}}
\definecolor{newcolor}{rgb}{.8,.349,.1}

\ifCLASSINFOpdf
\else
\fi
\begin{document}
%
\title{Evaluation of Various Open-Set Medical Imaging Tasks with Deep Neural Networks}
%
%
%

\author{Zongyuan Ge,~\IEEEmembership{Senior Member,~IEEE,}
        and Xin Wang,~\IEEEmembership{Member,~IEEE}
\thanks{Z. Ge is with eResearch centre, Monash University, 15 innovation walk, Clayton, VIC 3800, Australia  e-mail: zongyuan.ge@monash.edu}
\thanks{X. Wang is with Airdoc Research, 15 innovation walk, Clayton, VIC 3800, Australia}}

%
%

\markboth{ARXIV}%
{Shell \MakeLowercase{\textit{et al.}}: Bare Demo of IEEEtran.cls for IEEE Journals}
%



\maketitle

\begin{abstract}
The current generation of deep neural networks has achieved close-to-human results on ``closed-set'' image recognition; that is, the classes being evaluated overlap with the training classes. Many recent methods attempt to address the importance of the unknown, which are termed ``open-set'' recognition algorithms, try to reject unknown classes as well as maintain high recognition accuracy on known classes.
However, it is still unclear how different general domain-trained open-set methods from ImageNet would perform on a different but more specific domain, such as the medical domain.
Without principled and formal evaluations to measure the effectiveness of those general open-set methods, artificial intelligence (AI)-based medical diagnostics would experience ineffective adoption and increased risks of bad decision making.
In this paper, we conduct rigorous evaluations amongst state-of-the-art open-set methods, exploring different open-set scenarios from ``similar-domain'' to ``different-domain'' scenarios and comparing them on various general and medical domain datasets.
We summarise the results and core ideas and explain how the models react to various degrees of openness and different distributions of open classes. We show the main difference between general domain-trained and medical domain-trained open-set models with our quantitative and qualitative analysis of the results. We also identify aspects of model robustness in real clinical workflow usage according to confidence calibration and the inference efficiency.
Prior to our work, no research has been conducted to fully evaluate and leverage the effectiveness of open-set methods in the medical domain. Although many results show that current techniques are insufficient to address unknown inputs, they present new research opportunities to develop and improve open-set methods for practical medical AI applications, which can significantly benefit machine-assisted medical decisions in this critical domain.
The results in this paper are obtained from approximately 360 sets of experimental results over a total of 8,000 GPU hours on a machine equipped with GTX 1080Ti. The source code and models  needed to reproduce the experiments in this work will be made publicly available on our GitHub page.
\end{abstract}


\section{Introduction}\label{sec:intro}

\begin{figure*}[!t]
	\centering
	\includegraphics[scale=0.35]{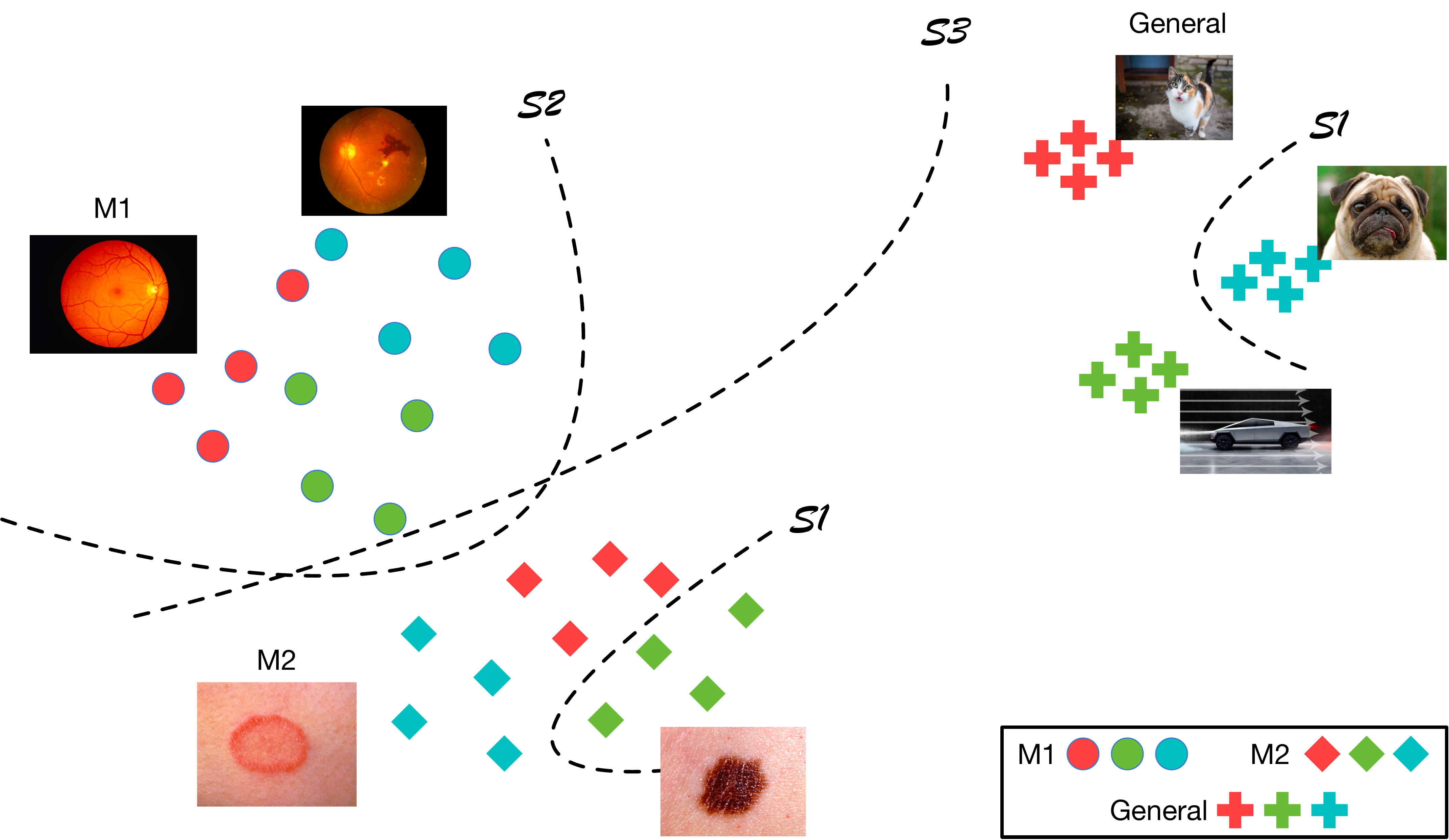}
\caption{This figure shows the blueprint for open-set evaluation tasks in this work. The shapes  denote various unique domain datasets. For instance, the circles and diamonds denote ophthalmology and dermatology domain images, respectively, while plus symbols denote the general image domain. The colours indicate various classes/conditions in each domain. General domain classes are relatively easy to cluster together since they share similar unique features (shape, colour, texture, etc.) among the same class, and those features are useful to distinguish them from other classes. Fine-grained imaging classification tasks, such as medical imaging diagnosis, are challenging since they have large intraclass variability (samples from the same condition have various appearances) and small interclass variability (visual disparity margins between various sub-conditions are quite similar). $S_{1},...,S_{3}$ denotes three open-set scenarios as described in Sec.~\ref{sec:sce} and Table.~\ref{tab:list-scenarios}.}
	\label{pic-overview}
\end{figure*}

Most existing deep learning-based clinical decision support models involve the ``\textit{closed-world assumption}'', where medical conditions during the test phase are assumed to overlap with the training conditions. In practice, a deployed medical diagnosis system must frequently face unseen conditions.
For example, there are more than 600 identified diseases of the nervous system, with causes ranging from genetic disorders and congenital abnormalities to infections and lifestyle factors. In general, there are unique challenges and burdens associated with each disease~\cite{world2006neurological}. Many of these disease are chronic, if not lifelong, conditions affecting people during childhood and their prime years.
Regarding practical model training, there are hundreds of skin conditions~\cite{meer2010rook}, but the largest publicly available skin dataset~\cite{tschandl2018ham10000} covers only eight common conditions. However, acquiring sufficient training data for rare conditions is as challenging as obtaining data on all possible diagnoses. With a clinical decision support system, even a small fraction of errors on unknowns could lead to disastrous outcomes.

A more feasible scenario for medical diagnosis applications is to assume an open set of conditions. We assume that the model trained with clinical knowledge is incomplete, and we expect that unseen medical conditions are encountered during the inference stage. From the learning perspective (see Fig.~\ref{pic-overview}), the medical diagnosis model trained from the centralised data should recognise known clinical cases accurately and deal with unknown conditions from the open space effectively.

In this study, we analyse and empirically clarify these issues, conducting a large set of rigorous experiments, starting with general class open-set recognition~\cite{scheirer2012toward} and continuing to the open-set medical diagnosis domain.
We consider \textbf{three scenarios}: deep learning-based open-set recognition methods on \textit{self-domain},  \textit{similar-domain}, and \textit{different-domain} classification; see Table~\ref{tab:list-scenarios} for details.
Our benchmark is verified on three common medical imaging diagnosis problems in dermatology, ophthalmology and radiology.

In this paper, we make the following contributions:
\begin{enumerate}
\item We fully explore the open-set problem on various medical domain tasks and investigate how open-set classification and uncertainty measurement methods designed for general images, where a domain shift exists, would perform on medical imaging diagnosis tasks.




\item We adopt a temperature scaling method to calibrate probabilities from the network to address the issue of uncalibrated probabilities of medical open-set disease recognition tasks.


\item We conduct experiments on several medical datasets and evaluate the feasibility (inference speed, etc.) of the methods in piratical clinical settings.


\item The source code, intermediate results and models to reproduce all the experiments in this work will be available on the project website for further benchmarking in the medical diagnosis field.

\end{enumerate}


%
%
%

%
%

\section{Related Work}\label{sec:related}
With the abundance of publicly available datasets and computational resources, deep learning, a subfield of machine learning, has been one of the most rapidly developed domains for the application of recognition machines~\cite{resnet,imageNet}. Computing-based medical imaging diagnosis enjoys great benefits from recent advances in object classification, detection and segmentation~\cite{ronneberger2015u,litjens2017survey}.
In the general imaging domain, the open-set problem has been a trending research topic in deep learning. \cite{bendale2016towards} proposed a new classification component called the OpenMax function combined with extreme value theory (EVT) to study the class score distribution of unknown classes. \cite{ge2017generative} extended the OpenMax approach to the generative OpenMax (G-OpenMAX) approach, which uses generative models to synthesise unknown samples and represent open space. A similar idea using counterfactual images is proposed by~\cite{neal2018open} to form a decision boundary between knowns and unknowns. \cite{oza2019c2ae} trained a class condition autoencoder (C2AE) model with sample-selective two-step training to acquire the open-set decision boundary. The open risk loss is constructed with the generative errors from the autoencoder.

Uncertainty learning for deep neural networks aims to measure the lack of knowledge from the prediction, which can be introduced into open-set recognition for unknown class detection. \cite{gal2016dropout} proposed a Monte Carlo (MC) dropout scheme to estimate the epistemic uncertainty in regression and classification tasks. \cite{lakshminarayanan2017simple} implemented an ensemble of Bayesian neural networks to yield predictive uncertainty estimates. However, there is a lack of literature and conclusions on how well uncertainty methods can be generalised to open-set detection problems.

In the medical domain, promising results from deep learning techniques have been demonstrated in the fields of ophthalmology~\cite{wu2019universal,wang2019retinal}, dermatology~\cite{nature,gu2019progressive}, radiology~\cite{sahiner2019deep,tao2019improving} and neurology~\cite{kuhlmann2018seizure}.
Although recent deep learning methods have advanced the fields of visual recognition and neural language understanding, we see a few challenges with the direct adaptation of deep learning to open-set recognition in medical tasks due to a lack of understanding of the models.
Our paper is the first to provide this important evaluation.


\begin{table*}[t!]

	\centering
		\caption{A list of scenarios.}
\begin{tabular}{| c | c | c | L |}\hline
\backslashbox{SCENARIOS}{SETTINGS}
&   KNOWN LABEL &   UNKNOWN LABEL & TARGET \\\hline

\textbf{Self-domain} & $\mathcal{L}_{known}^{M_{1}}$	&	$\mathcal{L}_{unknown}^{M_{1}}$ & {To recognise known and unknown conditions from the same medical domain} \\\hline

\textbf{Similar-domain}  & $\mathcal{L}_{known}^{M_{1}}$	& $\mathcal{L}^{M_{2}},\mathcal{L}^{M_{3}}$	& {To recognise known conditions from the same domain and unknown conditions from a similar medical domain} \\\hline

\textbf{Different-domain}  & $\mathcal{L}_{known}^{M_{1}}$	& $\mathcal{L}^{general}$	& {To recognise known conditions from the same medical domain and samples from the general imaging task} \\\hline

\end{tabular}
		\label{tab:list-scenarios}
\end{table*}

\begin{table}[!b]
	\centering
		\caption{A list of notations.}
	\begin{tabular}{cc}
		\hline
		\textbf{Notation}  & \textbf{Explanation}  \\ \hline
		$x$ & Image sample \\  
		$\hat y$  & Predicted condition \\ 
		$y$  & Disease condition \\
		$\phi(.)$       & Open-set condition function \\
		$f(.)$       & Recognition function \\
		$M$       & Medical domain task \\
		$\mathcal{L}_{known}$  & Known conditions from the database \\
		$\mathcal{L}_{unknown}$  & Unknown conditions in open space \\
		$\mathcal{U}$  & Unknown or open space \\
		$\mathcal{X}$  & Input space \\
		$\mathcal{Y}$  & Universe of all conditions \\
		\hline
		\end{tabular}
		\label{list-notation}
\end{table}

\section{Problem Formalisation}\label{sec:formal}
Here, we introduce the formalisation of open-set recognition with essential properties.
For clarity, a list of symbols is summarised in Table~\ref{list-notation}.
A classification model $f: \mathcal{X} \rightarrow \mathcal{L}_{known}$ denoting a mapping function for data samples in the input space $\mathcal{X}$ to a set of labels in $\mathcal{L}_{known}$ is trained during the training phase.
We define $\mathcal{L}_{unknown} = \mathcal{Y} \setminus  \mathcal{L}_{known}$ as the set of all unknown conditions that the model needs to reject during the inference phase.
Under the open-set condition, the assumption is that unknown medical conditions $\mathcal{L}_{unknown}$ do not appear in the training phase.
Being trained with only $\mathcal{L}_{known}$, an open-set classification model $f$ should have the ability to reject unknown samples as well as maintain a high classification accuracy on known classes.
The definition of the open-set learning problem is as follows:
\[
    \hat y = 
\begin{cases}
    Unknown, & \text{subject to } \phi(f(x))\\
    argmax_{y}P(y|x),              & \text{otherwise}
\end{cases}
\]
which develops mechanisms that are aware of and avoid misclassifying unseen conditions $\mathcal{L}_{unknown}$ with a reject option. The reject option is determined under the factor of open-set condition function $\phi$ (see Sec.~\ref{sec:method}). For example, $\phi$ can be a thresholding function of the class probability output $f(x)$.

In this work, we consider open-set medical diagnosis models trained without auxiliary or synthesised classes. Other open-set models trained with auxiliary data samples~\cite{panareda2017open} to approximate the unknown space $\mathcal{U}$ are not considered. The reason is that for the general image classification problem, for instance, the ImageNet challenge, many auxiliary data can be easily parsed from an internet image search engine or generated images from generative adversarial network (GAN) to represent the unknown space. However, many medical applications are restricted by both ethical and label noise (arising from a lack of expert verification) issues. Apart from that, many $\mathcal{L}_{unknown}$ can be rare and challenging to acquire. Synthesised samples from GAN-based models are not interpretable and do not maintain good semantic pathological meanings~\cite{xing2019adversarial}.





\section{Scenarios}\label{sec:sce}
This section introduces the three types of open-set recognition scenarios within the context of the medical domain.
Table~\ref{tab:list-scenarios} illustrates the key differences between all three scenarios.
To be consistent with the notations defined in Sec.~\ref{sec:formal}, different medical domain tasks (dermatology, radiology and ophthalmology, etc.) are defined as $\{ M_{1}, M_{2}, M_{3},...  \}$, and the corresponding known and unknown conditions for $M_{1}$, for example, are denoted by $\mathcal{L}_{known}^{M_{1}}$ and $\mathcal{L}_{unknown}^{M_{1}}$.
All the outlined open-set methods described in Sec.~\ref{sec:method} will be applied to all three of the following scenarios.

\subsection{Scenario 1: Self-domain open set}
We first explore the open-set scenario under the same medical disease domain. For example, a task to train a model to distinguish different skin conditions between various common benign and cancerous conditions (e.g., basal cell carcinoma (BCC), squamous cell carcinoma (SCC), and melanoma) and to reject other untrained rare skin conditions, such as parapsoriasis guttata.
Note that this scenario is also valuable for generating insights into the capabilities of the state-of-the-art general domain open-set methods on ``fine-grained'' medical conditions with large intraclass and small interclass variations~\cite{wei2019deep}.

This scenario is common in most medical AI applications. Using medical AI tools under a general practitioner's (GP's) guidance, the GP can help determine the medical imaging conditions without mistakes. However, without full knowledge of every pathology, even GPs are bounded by their subjectiveness and experiences in some rare and less-studied cases.
This limitation highlights the issue with a core assumption made in ``\textit{closed-set}'' scenarios that rare and unseen medical conditions are often observed in practice.


\subsection{Scenario 2: similar-domain open set}
In scenario 2, we explore the algorithm's performance when images from other medical imaging modalities are fed during inference. This is a more challenging scenario than the \textit{self-domain} scenario, as both medical domains $M_{1}$ and $M_{2}$ may share the similar property of small inter-class and large intra-class variations in the data distribution. Moreover, they are often consistent and close to each other in the colour space (dermoscopic images vs fundus images), which may confuse the model.

Scenario 2, that is, the \textit{similar-domain} open-set scenario, occurs when there are two confuseable domains. This phenomenon can occur when a clinician or operator makes a mistake in choosing the image modality. This circumstance may occur in less-developed medical institutions where data collected during routine care are stored on one central platform lacking management.




\subsection{Scenario 3: Different-domain open set}
In scenario 3, we explore the challenging setting in which general domain images and unknown conditions from other medical domains are fed together into the medical AI diagnosis system for model evaluation.
Unlike images from medical domains, general domain images (from ImageNet (see Sec.~\ref{sec:exp} for more details) are diverse in terms of colour and classes.
This scenario may occur in places where fully AI-supported screening services are provided through off-line and on-line health services without clinicians' supervision~\cite{wu2019universal,scheetz2019artificial}. Patients may upload irrelevant images by mistake or through carelessness.

\section{Method Details}\label{sec:method}
Here, we describe the recent deep learning-based open-set recognition methods evaluated in this work: OpenMax~\cite{bendale2016towards}, out-of-distribution detector for neural networks (ODIN)~\cite{liang2017enhancing}, MC dropout~\cite{gal2016dropout} and open long-tailed recognition (OLTR)~\cite{liu2019large}. They can be categorised into either Bayesian or non-Bayesian models.

\subsection{OpenMax}
OpenMax~\cite{bendale2016towards} takes advantage of both representation learning and EVT~\cite{rudd2017extreme} to tackle the open-set recognition problem.This method was one of the first works to explore deep representation learning from the penultimate layer to limit risk from open-space classes. First, a base network (AlexNet, residual neural network (ResNet), etc.) is trained as a penultimate activation layer extractor. EVT computes the points and distributions that best represent the property of each class. In the original paper, the mean activation vector (MAV) is employed to represent each class. It describes the functional form of the inclusion of a class with respect to another class. Then, a Weibull distribution is trained on top of known classes' post recognition activations to support the explicit rejection of unknown classes. Various distance-based loss functions, such as the centre loss~\cite{wen2016discriminative}, can be used along with the thresholding technique on top of the EVT outcomes.

\subsection{ODIN}
The ODIN framework~\cite{liang2017enhancing} was proposed to increase the difference between probability scores of known classes and unknown classes from the classifier layer of the neural network. It is a Bayesian approach for addressing the open-set classification problem. The ODIN framework has 
proven
to be effective on many open-set tasks~\cite{vyas2018out}.
It uses temperature calibration (often a constant quantity) on softmax scores. It shows that a calibration of the softmax output with large temperature values leads to more ``soft'' outputs, which push the unknowns class further away from the known classes. The unknown class tends to show a very different value of the largest logits compared to the known classes. Then, this property can be used to reject the unknown samples.

\subsection{MC Dropout}
\cite{miller2018dropout} shows that dropout variational inference is beneficial to open-set setting.
\cite{gal2016dropout} shows that a neural network with dropout applied before every layer is mathematically equivalent to the Bayesian model (probabilistic deep Gaussian process). This method is a computationally tractable approximation to obtain uncertainty estimates for the model. The implementation is to perform the sampling process by multiple stochastic forward passes with the dropout layers during inference and then average the softmax scores. For open-set recognition, like other open-set methods, we distinguish the known and unknown classes by thresholding the averaged score.


\subsection{OLTR}
The OLTR algorithm~\cite{liu2019large} was recently proposed to deal with real-world long-tailed and open-ended data distribution. It aims to train a feature space that embeds the closed-world classification while being able to detect the novelty of the open-set classes. Similar to OpenMax, the feature is directly obtained from the penultimate layer of a convolutional neural network (CNN), and the discriminative class centroid is created based on the distribution of the features. The concept of a memory module inspired by meta-learning~\cite{vinyals2016matching} is exploited to augment the features and thus enrich the tail class's embedding. The learned embedding functions are similar to EVT, as they are scaled inversely by their distance to the nearest class centroid. In other words, the further away a testing sample is from the class centroids, the more likely it is an open-set instance.


\section{Commonalities}\label{sec:common}
\begin{figure*}[!t]
	\centering
	\includegraphics[scale=0.32]{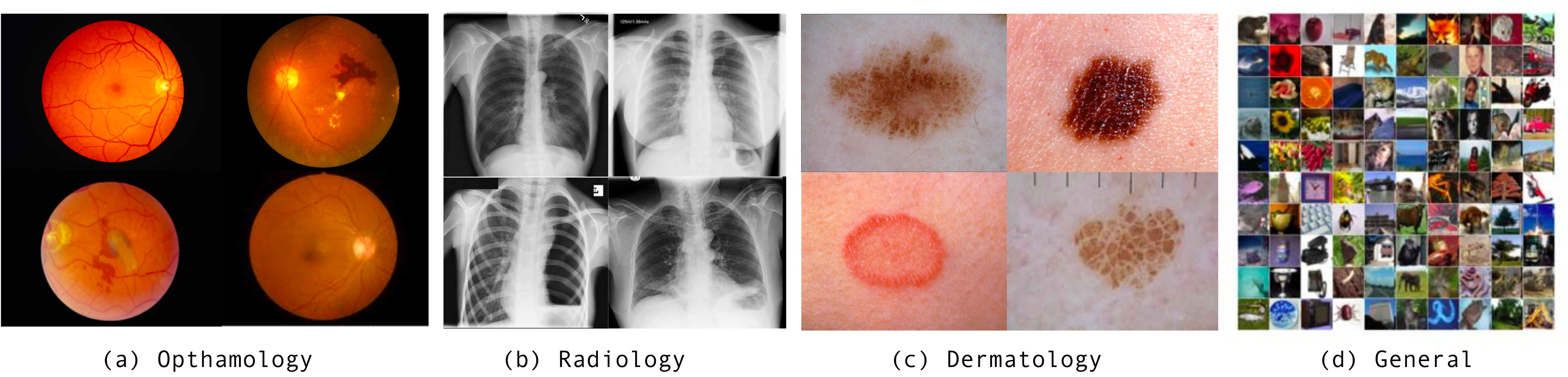}
\caption{An overview of the example images from three medical domains and one general domain as described in Sec.~\ref{sec:common}. As seen from (a), (b), and (c), various medical domain conditions can be defined as a fine-grained-like image classification problem because of small inter-class variations compared to the classes from general image datasets such as the Canadian Institute For Advanced Research (CIFAR) or ImageNet datasets in (d). It is noted that the general domain images differ significantly from most of the medical domain images in terms of texture, colour and shape. }
	\label{pic:domain diff}
\end{figure*}

In this section, we introduce what is common across the scenarios, datasets and evaluation metrics. Four datasets from different domains are presented with, and the selection criteria  are discussed. Sample images from three medical domains and one general domain are shown in Fig.~\ref{pic:domain diff}.

\subsection{Data}

\noindent\textbf{Dermatology (HAM10000):}
Approximately 5.4 million new skin cancers are diagnosed annually in the United States~\cite{rogers2015incidence}. Dermoscopic image-based computer-aided classification could be a critical tool for lesion tracking and early detection.
The Human Against Machine with 10000 training images (HAM10000) dataset consists of 10,015 dermoscopic images publicly available through
the International Skin Imaging Collaboration (ISIC) archive~\cite{HAM10000}. All the images were collected through digital dermoscopic images and non-digital diapositives.
The dermoscopic images ere acquired through digital dermoscopy with low levels of noise and consistent background illumination. Among all the images, approximately 50\% of lesions are confirmed by pathology. Biopsy is the gold standard in distinguishing malignant and benign lesions.
There are seven categories in HAM10000, consisting of five benign categories (i.e., melanocytic nevus, actinic keratosis, benign keratosis, dermatofibroma and vascular lesion) and two cancerous categories (i.e., melanoma and BCC).


\noindent\textbf{Ophthalmology (KaggleDR+):}
It is predicted that approximately 780 million people~\cite{pennington2016epidemiology} will develop retinal diseases such as diabetic retinopathy and age-related macular degeneration (AMD). Screening based on fundus imaging is a universally accepted strategy for vision-threatening disease referral, treatment and prevention.
To evaluate the open-set methods on fundus images, we use the relabelled multi-label retinal diseases dataset KaggleDR+ from~\cite{wang2019retinal}. The dataset consists of 18,614 re-labelled images from the Kaggle contest dataset~\footnote{https://www.kaggle.com/c/diabetic-retinopathy-detection}. Seventeen retinal diseases commonly examined during screening, such as diseases affecting the entire retina (diabetic retinopathy, etc.), optic disc (glaucoma, etc.) and macula (drusen, oedema, membranes, etc.). We used samples containing only a single disease label for consistent benchmarking with other datasets.


\noindent\textbf{Radiology (NIH Chest X-ray Dataset)}
Chest radiography is critical for the detection and triage of thoracic diseases such as tuberculosis and lung cancer.
Chest X-ray exams are one of the most common and cost-effective medical imaging screenings for various lung diseases, which affect millions of people worldwide each year~\cite{sahiner2019deep}.
ChestX-ray14~\cite{wang2019chestx} is currently the largest public thoracic disease dataset and includes 112,120 frontal-view X-rays with 14 disease labels~\footnote{The 14 diseases are atelectasis, consolidation, infiltration, pneumothorax, oedema, emphysema, fibrosis, effusion, pneumonia, pleural thickening, cardiomegaly, nodule mass, and hernia} from 30,805 unique patients' computed tomography (CT) slices. The distribution of the conditions is based on the frequency of diagnosis and observation in clinical practice.

\noindent\textbf{General domain (CIFAR-100)}
Most open-set recognition algorithms are designed and evaluated on some general imaging datasets. Here, we use CIFAR-100 as the general imaging dataset to reproduce benchmarks of various open-set algorithms.
The CIFAR-100 dataset~\cite{krizhevsky2009cifar} is a small general image dataset widely used to train computer vision and machine learning algorithms. It contains 100 different classes, such as aquatic mammals and household electrical devices.
It has 60,000 images with 600 samples of each class. In this work, to increase the difficulty of the open-set setting, our evaluation is performed on the 20 superclasses.

\subsection{Evaluation Metrics}
In this section, we describe the evaluation metrics used for close-set and open-set performance.

\noindent\textbf{FPR at 95\% TPR:} The false-positive rate (FPR) measures the fraction of unseen conditions $\mathcal{L}_{unknown}$ from open space misclassified as one of the known conditions $\mathcal{L}_{data}$ when the true-positive rate (TPR) is as high as 95\%. A high FPR indicates that unknown conditions are conflated with a known class.

\noindent\textbf{AUROC:} 
The receiver operating characteristic (ROC) curve describes the threshold-dependent relationship between the FPR and TPR, and the area under the ROC curve (AUROC) is a metric to evaluate the model performance trade-off between known conditions and unknown conditions. Here, all the known conditions are treated as positives, while the unknown conditions or open classes are treated as negatives. A 100\% AUCROC means a perfect detection of the known classes compared to the unknown classes.

\noindent\textbf{Detection Error:} The detection error is defined as $0.5(1-\text{TPR)} + 0.5\text{FPR}$ under the assumption that known and unknown conditions are equally distributed in the test set.

\noindent\textbf{F1 Score:} In this open-set task, we follow the modified F1 score from~\cite{geng2018recent,junior2017nearest} with the macro F-measure in Eq.~\ref{equ:macro}. The micro F-measure it does not treat unknown classes as one additional simple class because representative samples of unknown classes are not present during the training stage.
\begin{equation}
\small
P_{ma} = \frac{1}{C}\sum_{i=1}^{C}\frac{TP_{i}}{TP_{i}+FP_{i}}/C, R_{ma} = \frac{1}{C}\sum_{i=1}^{C}\frac{TP_{i}}{TP_{i}+FN_{i}}/C
\label{equ:macro}
\end{equation}

\begin{equation}
\small
P_{mi} = \frac{1}{C}\frac{\sum_{i=1}^{C}TP_{i}}{\sum_{i=1}^{C}(TP_{i}+FP_{i})}, R_{mi} = \frac{1}{C}\frac{\sum_{i=1}^{C}TP_{i}}{\sum_{i=1}^{C}(TP_{i}+FN_{i})}
\label{equ:macro}
\end{equation}
where $C$ denotes the number of known conditions. The macro-average computes the metric independently for each class and then takes the average, and hence, all the classes are treated equally); the micro-average aggregates the contributions of all the classes to compute the average metric.



\section{Experiments}\label{sec:exp}

\begin{table*}[t!]
\small
\caption{This table shows the known and unknown open-set classification performance in the self-domain scenario. } 
\begin{tabular}{cclllll}
\hline
\multicolumn{7}{c}{\textbf{Self-domain scenario}}                                                                                                                                                                                                                                                                                              \\ \hline
\textbf{Dataset}                                                                                                  & \textbf{Method} & \multicolumn{1}{c}{\textbf{FPR95 $\downarrow$}} & \multicolumn{1}{c}{\textbf{AUROC $\uparrow$}} & \multicolumn{1}{c}{\textbf{Detection $\downarrow$}} & \multicolumn{1}{c}{\textbf{F\_macro $\uparrow$}} & \multicolumn{1}{c}{\textbf{F\_micro $\uparrow$}} \\ \hline
\multicolumn{1}{c|}{\multirow{4}{*}{\begin{tabular}[c]{@{}c@{}}CIFAR-100\\ (2/5/10)\end{tabular}}}                & OpenMax         & \textbf{10.0} / \textbf{31.0} / \textbf{69.4}                 & 97.2 / 92.1 / 76.5                                   &  \textbf{6.7} / \textbf{16.0} / 28.0                                      &      \textbf{95.9} / \textbf{87.4} / \textbf{68.5}                                 &      \textbf{94.7} / \textbf{85.7} / \textbf{64.4}                                 \\
\multicolumn{1}{c|}{}                                                                                             & MC Dropout      & 54.9 / 79.7 / 96.4                                   &  82.7 / 74.2 / 64.6                                  &  19.9 / 30.5 / 37.5                                      &         92.0 / 82.6 / 59.4                              &       91.3 / 81.4 / 56.1                                \\
\multicolumn{1}{c|}{}                                                                                             & ODIN            & 26.5 / 31.6 / 78.3                & \textbf{97.8} / \textbf{97.2} / 72.6                 & 16.0 / 18.5 / 32.9                     & 95.3 / 86.6 / 66.5                    & \textbf{94.7} / \textbf{85.7} / 63.5                    \\
\multicolumn{1}{c|}{}                                                                                             & OLTR            & 57.9 / 48.4 / 78.8                 & 90.3 / 84.4 / \textbf{79.3}                 & 14.9 / 20.0 / \textbf{23.1}                     & 95.2 / 85.5 / 67.3                    & \textbf{94.7} / 85.1 / 63.4                    \\ \hline
\multicolumn{1}{c|}{\multirow{4}{*}{\begin{tabular}[c]{@{}c@{}}Dermatology\\ ISIC-2019\\ (1/2/3)\end{tabular}}}   & OpenMax         &  \textbf{66.8} / \textbf{82.5} / \textbf{82.1}                                  &   \textbf{84.8} / 74.2 / \textbf{71.2}                                 &   \textbf{22.9} / \textbf{34.3} / \textbf{34.2}                                    &        72.2 / 78.3 / 58.5                               &        80.3 / 83.1 / 54.0                               \\
\multicolumn{1}{c|}{}                                                                                             & MC Dropout      &     82.4 / 93.6 / 96.9                               &      67.5 / 61.2 / 45.8                              &   36.7 / 41.0 / 48.9                                     &    41.9 / 52.2 / 43.4                                   &      60.9 / 69.0 / 39.5                                 \\
\multicolumn{1}{c|}{}                                                                                             & ODIN            & 74.8 / 83.1 / 88.4                 & 83.1 / \textbf{76.7} / 69.7                & 24.7 / 34.5 / 35.5                    & 72.0 / \textbf{78.3} / 58.8                    & \textbf{80.6} / \textbf{83.1} / 53.9                    \\
\multicolumn{1}{c|}{}                                                                                             & OLTR            & 84.0 / 86.7 / 91.4                 & 73.8 / 69.3 / 61.6                 & 30.3 / 34.8 / 40.5                     & \textbf{72.9} / 78.2 / \textbf{67.2}                    & 79.6 / 81.8 / \textbf{55.2}                    \\ \hline
\multicolumn{1}{c|}{\multirow{4}{*}{\begin{tabular}[c]{@{}c@{}}Radiology\\ NIH14\\ (4/6/8)\end{tabular}}}         & OpenMax         &      89.8 / 93.4 / \textbf{92.3}                             &    59.3 / \textbf{58.1} / 55.6                                &   43.0 / \textbf{43.3} / \textbf{45.4}                                     &    23.8 / 27.5 / \textbf{35.6}                                   &       44.1 / 47.0 / \textbf{47.4}                                \\
\multicolumn{1}{c|}{}                                                                                             & MC Dropout      &          \textbf{87.9} / 94.3 / 94.6                          &       \textbf{59.9} / 57.1 / 53.2                             &     \textbf{42.5} / 44.1 / 46.7                                   &         26.5 / \textbf{27.8} / 7.7                              &        \textbf{45.1} / \textbf{47.2} / 38.0                               \\
\multicolumn{1}{c|}{}                                                                                             & ODIN            & 92.0 / \textbf{91.8} / 92.5                 & 57.8 / 57.2 / \textbf{56.4}                 & 43.7 / 44.6 / 45.5                     & 25.2 / 27.2 / 35.5                    & 43.9 / 46.2 / 46.9                    \\
\multicolumn{1}{c|}{}                                                                                             & OLTR            & 94.8 / 93.4 / 94.3                 & 51.3 / 55.8 / 53.0                 & 48.7 / 45.6 / 47.2                     & \textbf{28.5} / 26.9 / 35.4                    & 36.2 / 43.8 / 44.8                    \\ \hline
\multicolumn{1}{c|}{\multirow{4}{*}{\begin{tabular}[c]{@{}c@{}}Ophthalmology\\ KaggleDR+\\ (2/5/9)\end{tabular}}} & OpenMax         &     \textbf{59.2} / \textbf{79.4} / \textbf{77.1}                               &      77.2 / \textbf{71.6} / 62.5                              &     28.3 / \textbf{34.5} / 39.7                                  &      52.6 / \textbf{64.0} / 60.1                                 &      \textbf{68.5} / 69.4 / \textbf{67.2}                                 \\
\multicolumn{1}{c|}{}                                                                                             & MC Dropout      &        70.1 / 95.9 / 85.3                            &         \textbf{78.9} / 53.7 / \textbf{65.0}                           &      \textbf{28.1} / 45.9 / \textbf{38.2}                                  &            29.0 / 29.9 / 38.2                           &    45.5 / 54.4 / 47.8                                   \\
\multicolumn{1}{c|}{}                                                                                             & ODIN            & 84.3 / 88.1 / 90.4                 & 65.2 / 69.7 / 61.4                 & 38.6 / 36.4 / 44.2                     & 56.8 / 63.4 / 56.6                    & \textbf{68.9} / \textbf{69.6} / 66.9                   \\
\multicolumn{1}{c|}{}                                                                                             & OLTR            & 90.3 / 87.6 / 94.0                 & \textbf{45.2} / 66.9 / 42.6                 & 47.2 / 37.2 / 49.0                     & \textbf{57.9} / 58.2 / \textbf{62.0}                    & 66.7 / 67.3 / 66.7                    \\ \hline
\end{tabular}

\label{tab:self-domain}
\end{table*}

\begin{table*}
\small
\caption{This table shows the known and unknown open-set classification performance in the similar-domain scenario. } 
\begin{tabular}{cclllll}
\hline
\multicolumn{7}{c}{\textbf{Similar-domain scenario}}                                                                                                                                                                                                                                                                                            \\ \hline
\textbf{Dataset}                                                                                                  & \textbf{Method} & \multicolumn{1}{c}{\textbf{FPR95 $\downarrow$}} & \multicolumn{1}{c}{\textbf{AUROC $\uparrow$}} & \multicolumn{1}{c}{\textbf{Detection $\downarrow$}} & \multicolumn{1}{c}{\textbf{F\_macro $\uparrow$}} & \multicolumn{1}{c}{\textbf{F\_micro $\uparrow$}} \\ \hline
\multicolumn{1}{c|}{\multirow{4}{*}{\begin{tabular}[c]{@{}c@{}}Dermatology\\ ISIC-2019\\ (1/2/3)\end{tabular}}}   & OpenMax         & \textbf{81.5} / 74.3 / 79.5                 & \textbf{79.8} / 83.1 / 77.1                 & 26.6 / 23.9 / 28.4                     & \textbf{78.4} / 79.7 / 55.1                    & 80.3 / 83.1 / 54.0                    \\
\multicolumn{1}{c|}{}                                                                                             & MC Dropout      & 91.3 / 95.3 / 90.2                 & 63.3 / 69.3 / 71.9                 & 40.1 / 29.6 / 32.4                     & 72.3 / 76.5 / 61.3                    & 77.2 / 80.7 / 50.6                    \\
\multicolumn{1}{c|}{}                                                                                             & ODIN            & 84.5 / 80.5 / 82.0                 & 79.2 / 82.4 / 81.3                 & 27.0 / 25.1 / 24.4                     & 78.0 / \textbf{80.5} / 55.0                    & \textbf{80.3} / \textbf{83.1} / 53.9                    \\
\multicolumn{1}{c|}{}                                                                                             & OLTR            & 81.6 / \textbf{58.9} / \textbf{75.6}                 & 79.0 / \textbf{91.1} / \textbf{84.2}                 & \textbf{26.0} / \textbf{14.3} / \textbf{21.6}                     & 75.8 / 74.7 / \textbf{66.6}                    & 79.4 / 81.7 / \textbf{55.5}                    \\ \hline
\multicolumn{1}{c|}{\multirow{4}{*}{\begin{tabular}[c]{@{}c@{}}Radiology\\ NIH14\\ (4/6/8)\end{tabular}}}         & OpenMax         & 100.0 / 99.6 / 99.8                 & 16.9 / 15.3 / 26.5                 & 50.0 / 50.0 / 50.0                     & 28.3 / 30.8 / \textbf{42.1}                      & 43.0 / 46.2 / \textbf{47.3}                    \\
\multicolumn{1}{c|}{}                                                                                             & MC Dropout      & 99.7 / 83.7 / 89.8                 & 26.0 / \textbf{85.1} / 58.5                 & 50.0 / \textbf{18.7} / 40.0                     & 28.1 / 27.0 / 39.1                       & \textbf{45.2} / \textbf{47.0} / 45.7                    \\
\multicolumn{1}{c|}{}                                                                                             & ODIN            & 100.0 / 99.8 / 99.4                  & 16.3 / 16.8 / 34.2                 & 49.9 / 50.0 / 50.0                     & 28.6 / \textbf{31.4} / 42.0                    & 43.4 / 46.5 / 47.0                    \\
\multicolumn{1}{c|}{}                                                                                             & OLTR            & \textbf{89.3} / \textbf{61.8} / \textbf{61.4}                 & \textbf{57.5} / 84.9 / \textbf{76.2}                 & \textbf{43.5} / 23.3 / \textbf{30.4}                     & \textbf{29.7} / 26.1 / 32.7                    & 35.7 / 43.1 / 44.8                    \\ \hline
\multicolumn{1}{c|}{\multirow{4}{*}{\begin{tabular}[c]{@{}c@{}}Ophthalmology\\ KaggleDR+\\ (2/5/9)\end{tabular}}} & OpenMax         & 45.7 / 55.9 / 59.0                 & \textbf{84.0} / 77.3 / 81.2                 & \textbf{22.2} / 23.9 / 26.7                     & 49.1 / \textbf{59.7} / \textbf{62.6}                    & \textbf{68.9} / 69.4 / \textbf{67.2}                    \\
\multicolumn{1}{c|}{}                                                                                             & MC Dropout      & \textbf{42.6} / \textbf{45.6} / \textbf{46.1}                 & 78.0 / 80.4 / 76.3                 & 23.3 / 23.7 / 24.9                     & 22.7 / 21.1 / 32.7                       & 45.6 / 54.4 / 47.9                    \\
\multicolumn{1}{c|}{}                                                                                             & ODIN            & 78.0 / 70.4 / 73.0                 & 71.6 / \textbf{82.0} / \textbf{82.2}                 & 33.6 / \textbf{23.2} / \textbf{24.2}                     & 52.7 / 59.4 / 62.3                    & 68.4 / \textbf{69.4} / 67.1                    \\
\multicolumn{1}{c|}{}                                                                                             & OLTR            & 78.4 / 68.0 / 80.1                 & 63.1 / 80.1 / 61.4                 & 38.1 / 25.2 / 38.6                     & \textbf{53.4} / 52.3 / 55.7                    & 66.8 / 66.5 / 65.7                    \\ \hline
\end{tabular}
\label{tab:similar-domain}
\end{table*}

\begin{table*}
\small
\caption{This table shows the known and unknown open-set classification performance in the different-domain scenario. } 
\begin{tabular}{cclllll}
\hline
\multicolumn{7}{c}{\textbf{Different-domain scenario}}                                                                                                                                                                                                                  \\ \hline
\textbf{Dataset}                                                                                                  & \textbf{Method} & \multicolumn{1}{c}{\textbf{FPR95 $\downarrow$}} & \multicolumn{1}{c}{\textbf{AUROC $\uparrow$}} & \multicolumn{1}{c}{\textbf{Detection $\downarrow$}} & \multicolumn{1}{c}{\textbf{F\_macro $\uparrow$}} & \multicolumn{1}{c}{\textbf{F\_micro $\uparrow$}} \\ \hline
\multicolumn{1}{c|}{\multirow{4}{*}{\begin{tabular}[c]{@{}c@{}}Dermatology\\ ISIC-2019\\ (1/2/3)\end{tabular}}}   & OpenMax         & 85.3 / 77.0 / 89.9                 & 77.3 / 84.2 / 63.0                 & 28.3 / 22.2 / 40.0                     & 78.2 / 78.8 / 65.7                    & 80.3 / 83.1 / 54.0                    \\
\multicolumn{1}{c|}{}                                                                                             & MC Dropout      & 96.7 / 96.2 / 99.1                 & 54.8 / 62.6 / 23.1                 & 45.1 / 39.2 / 50.0                     & 70.4 / 73.7 / 64.5                       & 76.9 / 80.7 / 50.5                   \\
\multicolumn{1}{c|}{}                                                                                             & ODIN            & \textbf{83.9} / \textbf{74.1} / 93.2                 & \textbf{79.0} / \textbf{86.7} / 63.1                 & \textbf{27.1} / \textbf{19.2} / 40.8                     & 78.3 / 78.9 / 64.1                    & \textbf{80.3} / \textbf{83.1} / 53.9                    \\
\multicolumn{1}{c|}{}                                                                                             & OLTR            & 85.9 / 81.5 / \textbf{85.5}                 & 77.1 / 84.3 / \textbf{77.1}                 & 28.3 / 20.1 / \textbf{27.6}                     & \textbf{79.7} / \textbf{79.7} / \textbf{69.3}                    & 79.7 / 81.6 / \textbf{55.8}                    \\ \hline
\multicolumn{1}{c|}{\multirow{4}{*}{\begin{tabular}[c]{@{}c@{}}Radiology\\ NIH14\\ (4/6/8)\end{tabular}}}         & OpenMax         & 99.9 / 98.6 / 98.7                 & 13.4 / 12.6 / 31.0                 & 50.0 / 50.0 / 50.0                    & 28.1 / 31.8 / \textbf{41.8}                     & 42.9 / 46.2 / \textbf{47.3}                    \\
\multicolumn{1}{c|}{}                                                                                             & MC Dropout      & \textbf{99.9} / 98.6 / 95.4                 & 37.2 / 66.5 / 59.5                 & 49.8 / 32.3 / 40.8                     & 29.4 / 28.6 /41.2                      & \textbf{45.2} / \textbf{47.0} / 45.9                    \\
\multicolumn{1}{c|}{}                                                                                             & ODIN            & 99.9 / 99.9 / 98.6                 & 11.9 / 11.3 / 37.1                 & 50.0 / 50.0 / 50.0                     & 28.5 / \textbf{32.3} / 41.2                    & 43.4 / 46.6 / 47.1                    \\
\multicolumn{1}{c|}{}                                                                                             & OLTR            & \textbf{96.9} / \textbf{67.3} / \textbf{75.1}                 & \textbf{55.6} / \textbf{80.3} / \textbf{72.8}                 & \textbf{40.0} / \textbf{27.2} / \textbf{34.1}                     & \textbf{30.4} / 27.8 / 31.3                    & 35.6 / 44.7 / 44.9                    \\ \hline
\multicolumn{1}{c|}{\multirow{4}{*}{\begin{tabular}[c]{@{}c@{}}Ophthalmology\\ KaggleDR+\\ (2/5/9)\end{tabular}}} & OpenMax         & 84.3 / \textbf{76.7} / \textbf{61.3}                 & 45.7 / 77.3 / \textbf{78.5}                 & 43.8 / 28.0 / \textbf{28.4}                     & 53.7 / 62.1 / \textbf{64.0}                    & \textbf{68.5} / 69.4 / 67.2                    \\
\multicolumn{1}{c|}{}                                                                                             & MC Dropout      & 71.7 / 93.2 / 72.9                 & \textbf{81.9} / 78.4 / 70.6                 & \textbf{23.7} / \textbf{22.0} / 33.1                     & 29.2 / 26.0 / 36.2                       & 45.6 / 54.5 / 47.9                    \\
\multicolumn{1}{c|}{}                                                                                             & ODIN            & 89.6 / 82.6 / 80.6                & 52.5 / 77.6 / 76.1                 & 45.3 / 27.7 / 30.1                     & 57.0 / \textbf{64.3} / 63.7                    & 68.4 / \textbf{69.4} / \textbf{67.2}                    \\
\multicolumn{1}{c|}{}                                                                                             & OLTR            & \textbf{63.7} / 78.8 / 81.3                 & 68.0 / \textbf{79.2} / 51.1                 & 34.1 / 27.0 / 43.0                     & \textbf{58.3} / 55.0 / 57.4                    & 66.8 / 66.6 / 66.3                    \\ \hline
\end{tabular}
\label{tab:different-domain}
\end{table*}

\subsection{Implementation details}
All the algorithm frameworks are implemented in Python based on the PyTorch library~\cite{ketkar2017introduction}. We used ResNet-51~\cite{he2016deep} as the backbone network with the weights trained on the ImageNet dataset and with Adam as the optimiser. Data augmentation is adopted to expand the training set by random flipping and cropping. All the optimised hyperparameter settings for various open-set methods can be found in the supplementary for reference and also at the project code repository~\footnote{https://github.com/zongyuange/openset-tmi19/tree/master/bash}. The optimal parameters are chosen to minimise the FPR at a TPR of 95\% on the validation set.

\noindent\textbf{Openness Score:} We evaluated each open-set algorithm on each domain with different openness scores as defined in~\cite{scheirer2012toward}.
\begin{equation}
openness  = 1 - \sqrt{\frac{2N_{Train}}{N_{R}+N_{Test}}}
\end{equation}
where $N_{Train}$ defines the number of training classes, $N_{Test}$ denotes the number of testing classes and $N_{R}$ is the number of classes to be recognised.
A large openness score means that more classes are allocated in the open space and towards more open problems. When the openness score equals zero, it indicates that the problem is a closed-set problem. Known and unknown classes are randomly selected from the open class/condition pool. A detailed class distribution can be found on the project page.

\noindent\textbf{Data Partitioning:} The sample set of each condition is divided into training and validation sets (70\% and 15\%, respectively) to find the optimal hyperparameters for OpenMax, ODIN, MC Dropout and OLTR. The remaining 15\% of the samples are used for testing.
Details of class distribution for each dataset can be found in supplementary materials. 
During evaluation of the three scenarios, we tend to keep the same total number of testing samples for the self-domain, similar-domain and different-domain evaluations.
The ratio between the \textit{closed-set} and \textit{open-set} scenarios is fixed during inference.

\subsection{Open-set results}
In Table~\ref{tab:self-domain}, Table~\ref{tab:similar-domain} and Table~\ref{tab:different-domain}, we show our experimental results on the three different scenarios with multiple metrics (listed in Sec.~\ref{sec:common}).
All the values are percentages; $\uparrow$ means that a larger value indicates a better performance, while $\downarrow$ means that a lower value indicates a better performance. The number of unknown classes/conditions corresponding to different openness scores for each dataset is listed under the dataset name.

\subsubsection{Open-set baseline on the general domain dataset}
In the upper part of Table~\ref{tab:self-domain}, we first report all the candidate open-set methods' performance on the CIFAR-100 images.
What stands out in these results is the large performance difference among the various methods.
ODIN outperforms the other methods in terms of the binary evaluation metric AUROC, while OpenMax demonstrates better results in multi-class setting according to F$\_$macro and F$\_$micro.
Further analysis shows that there is a significant negative correlation between the openness score and AUROC. What is surprising is that OpenMax and ODIN achieve quite a robust performance with an increasing number of open classes.
A possible explanation for this phenomenon might be that these two methods have a post-probability calibration mechanism (temperature scaling and Weibull distribution) to stabilise the prediction score so that it is not pulled too far away by the unknown classes.
The results of FPR95 and the detection error further confirm that OpenMax and ODIN are more robust against open-space risk than the other methods.
The Bayesian-based uncertainty measurement method MC dropout does not achieve a competitive performance in this setting, suggesting that there exists a hypothesis gap between the uncertainty approximation and open-set recognition tasks. We will explore this concept further in the following testing scenarios.



\subsubsection{\textbf{Scenario 1:} The self-domain open-set scenario}
\label{sec: s1}
We first evaluate the \textit{self-domain} performance with untrained same-domain medical conditions as open-set classes.
Overall, it is apparent from Table~\ref{tab:self-domain} that the medical domain tasks are more challenging than the general image domain task under the \textit{self-domain} scenario. For instance, the AUROC for the medical domain tasks (with the lowest openness score) ranged between $67.5\%$ to $84.8\%$ for dermatology, $51.3\%$ to $59.9\%$ for radiology and $45.2\%$ to $78.9\%$ for ophthalmology, compared to $90.3\%$ to $97.8\%$ in the general image domain task.
Similar trends can also be observed at different openness score settings on close-set class metrics F$\_$macro and F$\_$micro.

A further observation of the dataset complexity among the three medical domain tasks from Table~\ref{tab:self-domain} is that most methods perform better on the dermatology dataset and worse on the NIH14 radiology dataset.
The poor performance on the radiology task can be explained by two main factors. First, most NIH14 labels are directly extracted from the radiology report without further verification~\cite{luke14}. Second, instead of having predictable pathology locations such as ophthalmology tasks, where pathologies such as glaucoma and AMD tend to appear only in the macular region, conditions such as pneumonia or atelectasis from the NIH14 dataset can appear in any regions in the lungs (see Fig.~\ref{pic:domain diff}), which makes this task more challenging than the other two.



Overall, the OpenMax approach performs consistently well on dermatology and ophthalmology tasks and is often the best performing approach on open-set risk metrics such as FPR95 and detection and competitive results on F$\_$macro and F$\_$micro. If we now turn to AUROC from ISIC-2019, ODIN is the second-best approach on this task where the uncertainty-based method MC dropout performs the worst. Contrary to expectations, AUROC on KaggleDR+ indicates that MC dropout has competitive performance when the number of conditions is large. However, its robustness and stability are questionable owing to contradictory results on AUROC and F$\_$macro with various openness scores.


\subsubsection{\textbf{Scenario 2:} The similar-domain open-set scenario} In the second scenario, we empirically evaluate how \textit{similar-domain} conditions from other medical domains  affect the open-set methods' performance. When training the dermatology model as the source model, we randomly sample an equal number of \textit{similar-domain} images from the radiology and ophthalmology datasets as open-set samples.


The results are shown in Table~\ref{tab:similar-domain}. When looking at the overall AUROC performance across three medical domain tasks, we can observe that open-set methods predict unknown open conditions better on \textit{similar-domain} than \textit{self-domain} on dermatology and ophthalmology tasks.
For instance, the top-performing OLTR method achieves $79.0\%$ to $91.1\%$ on \textit{similar-domain} for dermatology compared to $61.6\%$ to $73.8\%$ under the \textit{self-domain} setting on the exact same known condition distribution.
These results are likely to be related to imaging samples from different medical domains incorporating different morphological appearances, which makes the trained model reject out-of-distribution conditions more easily.

Contrary to expectations, most methods using the ophthalmology model as the source more easily recognise open samples than when using the dermatology model in the \textit{similar-domain} scenario. The AUROCs for dermatology and ophthalmology with the OpenMax, MC dropout and ODIN methods support this conclusion. Another interesting finding is that all the methods except OLTR fail on the NIH14 radiology dataset with worse results than the corresponding ones in the \textit{self-domain} scenario.
This discrepancy could be attributed to a similar reason, which was described in Sec.~\ref{sec: s1}. Another hypothesis is that although X-ray images look very different from natural images in the pixel space, the representation space learned from the radiology dataset does not encode the colour information because it is not useful in distinguishing various lung diseases.


Regarding the methods, OpenMax is no longer the top performer in this scenario. Instead, OLTR consistently outperforms ODIN on the ISIC-2019 datasets and even on the noisy NIH14 dataset by a large margin. This finding could be attributed to OLTR easily acknowledging the novelty of the open world by referencing the visual concepts dictionary.
One unanticipated finding is that the MC dropout method is the champion in terms of the FPR95 score in the ophthalmology task, whereas it performs badly in terms of F$\_$macro.
These observations indicate that open-set methods are task-dependent and dataset-dependent.



\begin{figure*}[!t]
	\centering
	\includegraphics[scale=0.4]{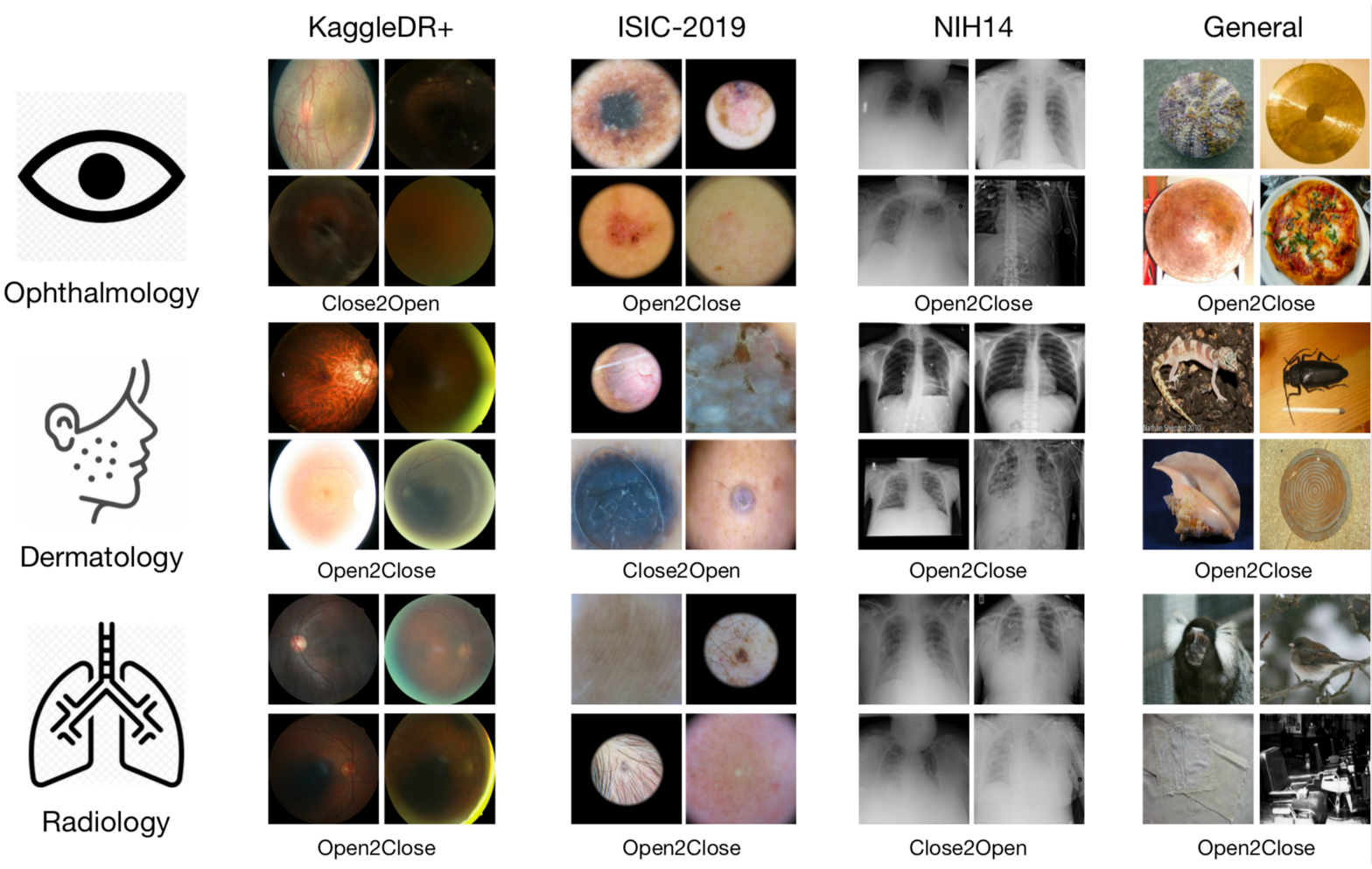}
\caption{This figure shows qualitative results on the various datasets. The leftmost symbols denote the ``closed-set'' trained source domain, and the corresponding results from the KaggleDR+, ISIC-2019, NIH14 and ImageNet datasets are shown on the right side of the figure. Close2Open represents false negatives (FNs), while Open2Close represents false positives (FPs).}
	\label{pic-qua}
\end{figure*}

\subsubsection{\textbf{Scenario 3:} The different-domain open-set scenario} To maintain the consistency of the input image dimension with ISIC-2019, NIH14 and KaggleDR+ and to increase the diversity to represent real open-set classes from the general domain, we supply open space samples for the general domain from the Tiny ImageNet dataset rather than CIFAR-100. Tiny ImageNet~\footnote{https://tiny-imagenet.herokuapp.com/} has 120,000 images covering 200 classes.

In Table~\ref{tab:different-domain}, we report our results for scenario 3. We first compare the overall performance to the \textit{self-domain} and \textit{similar-domain} results.
First, we note that the performance metrics for the OpenMax, ODIN and OLTR methods such as FPR95 and AUROC on the ISIC-2019 and KaggleDR+ datasets are worse than those reported in the \textit{similar-domain} scenario but better than those reported in the \textit{self-domain} scenario. This finding demonstrates that general domain images are more difficult to distinguish than medical domain images as open-space samples due to their large class and representation diversity. However, similar performance can be observed in terms of F$\_$macro and F$\_$micro across all the models in this scenario when compared to the results in the \textit{similar-domain} scenario.
This finding indicates that those open-set models have good consistency in recognising known classes that do not depend on the types of open-set conditions.

Regarding the methods on various scenarios, ODIN and OLTR are the better two models and are more robust when rejecting open classes when using dermatology as the source mode because the OLTR method's mechanism of building the class centroids to minimise the adverse effects from unbalanced learning and open space makes it achieve a competitive performance on a few challenging open-set settings. This mechanism also explains its low FPR95 value on the noisy NIH14 dataset.
ODIN with a high temperature value is able to put high ``soft'' margins on samples from the general domain.
OpenMax performs well on recognising general domain images when trained with KaggleDR+, a dataset that provides more diversity than the others in terms of the number of conditions, as diversity for training conditions can lead to a discriminative EVT model.



\begin{figure*}[!t]
	\centering
	\includegraphics[scale=0.5]{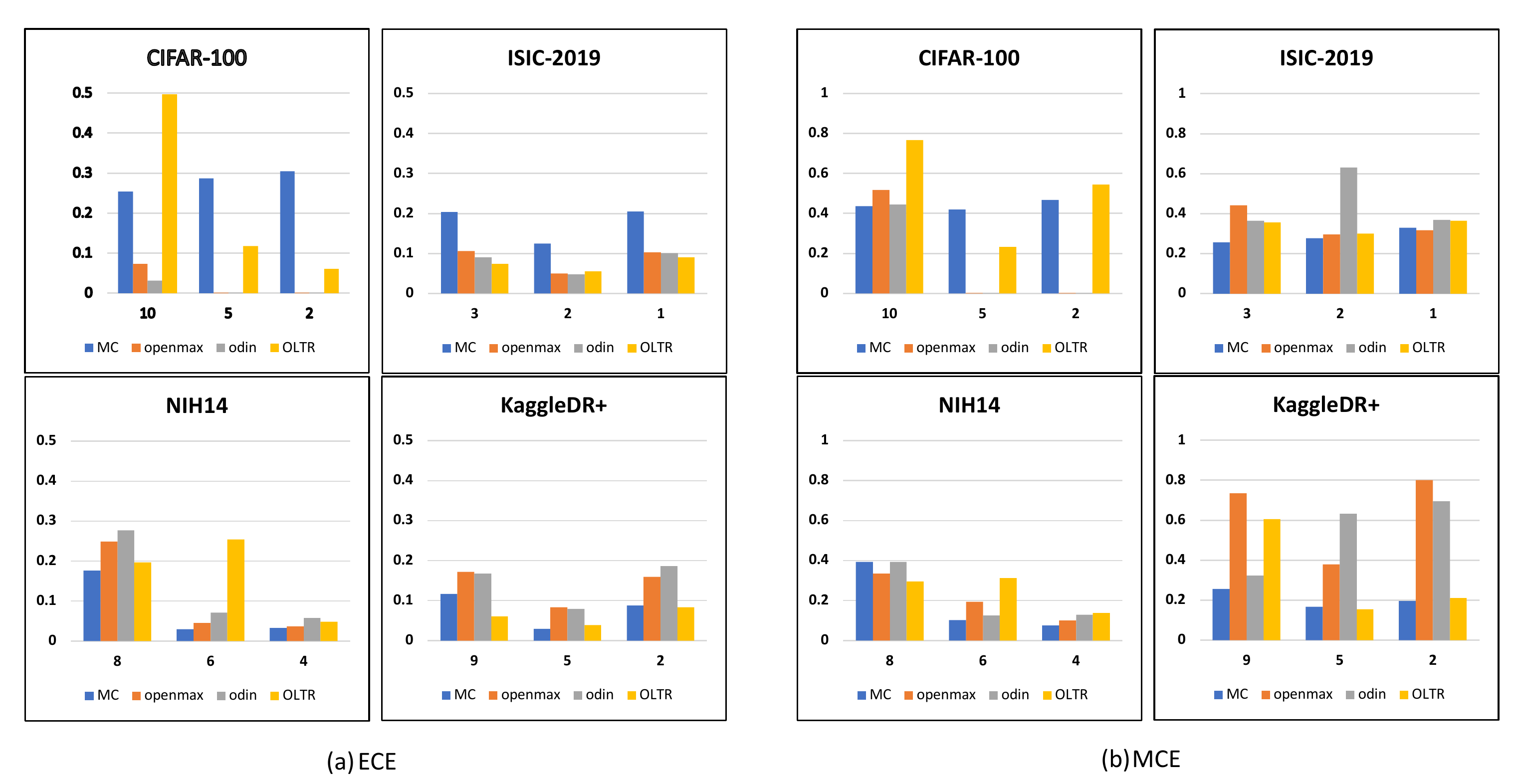}
\caption{This figure shows the ECE and MCE for the open-set method calibration evaluation in this work. The horizontal axis denotes the openness score, the vertical axis denotes the error scale, and the depth axis represents the various open-set methods.}
	\label{pic-chart}
\end{figure*}

\subsection{Qualitative analysis}
In this subsection, we show the qualitative results of open-set methods on four datasets in Fig.~\ref{pic-qua}. Sample images are selected using OLTR results as it shows relatively robust performance across most scenarios. We mainly present two types of observations, Open2Close (false positives), where unknowns are recognised as known classes, and Close2Open (false negatives), where known classes are rejected as out-of-distribution samples.
For the ophthalmology images in the first row, some circle-shaped objects are confused as retinal images. On the other hand, some low-quality (blurry and low-intensity) retinal images are treated as outliers. Similar trends can be observed when the dermatology and radiology images are trained as the source domain. In contrast, the dermatology model fails to reject samples from the KaggleDR+ and ImageNet datasets, where objects are presented in the centre with relatively uniform backgrounds. This finding demonstrates that the model preserves the prior knowledge where lesions tend to stay in the centre in most dermoscopic images~\cite{gu2019progressive}.
Finally, we observe samples in grey from the general domain, and images lacking texture information from the medical datasets can easily confuse the radiology model.

\subsection{Model calibration results}
\label{sec:calibra}
The prediction of the neural network probability representative of the true correctness likelihood is important for many decision-making pipelines of AI applications, such as self-driving vehicles or medical diagnosis. The trustworthiness of the calibrated confidence could also improve model interpretability and reliability. Here, we evaluate four open-set methods' confidence calibration capability on the ``closed set''~\footnote{The evaluated models for calibration are optimised in terms of the parameter settings on the open-set validation set}.
The two metrics used in these experiments are the expected calibration error (ECE) and the maximum calibration error (MCE), as proposed by~\cite{guo2017calibration}. The calibration error results of the open-set methods on these four datasets are shown in Fig.~\ref{pic-chart}. From the figure, we observe that most models and datasets experience some degree of miscalibration. The NIH14 dataset yields a higher average ECE than the other datasets but a relatively lower MCE. This finding means that the worst-case deviation between confidence and accuracy is much higher on the CIFAR-100, ISIC-2019 and KaggleDR+ datasets.

From the perspective of model calibration performance, the most important discovery is that one of the best calibration temperature scaling methods, ODIN~\cite{guo2017calibration}, does not always outperform the other non-calibration-oriented methods. In other words, open-set boundary and confidence calibration do not share the same optimisation goal, and they perhaps counteract each other's effects. OLTR performs well on datasets with diverse conditions (i.e., ISIC-2019 and KaggleDR+) and fails to produce competitive results on the NIH14 dataset. The Bayesian method, MC dropout, demonstrates good calibration results on the less challenging datasets such as KaggleDR+ and best on the most challenging NIH14 dataset when openness score is high, although it is not good at rejecting open-set samples.

\section{Discussion and Future Directions}
In this subsection, we summarise and analyse the effectiveness and robustness of the open-set model on various medical domain tasks.

\noindent\textbf{General domain vs medical domain open-set recognition:}
Through the analysis from Sec.~\ref{sec:exp}, we observe that open-set recognition for medical domain tasks is more challenging than that for general domain tasks according to the performance drop across the medical datasets. Despite the fact that various applications in general open-set recognition have been proposed, these methods are considered unreliable in fine-grained medical domain tasks. Our study shows the necessity to reconsider model design when dealing with fine-grained medical domain in-distribution and out-of-distribution conditions.

\noindent\textbf{Model generalisability:}
We then discuss the model generalisability with open conditions from various sources.
First, we notice that the ODIN, MC dropout, OpenMax and OLTR models are not trained from scratch using the source medical domain data. All the models use the pre-trained ImageNet weights due to the lack of large and high-quality publicly available medical data.
When open-set classes from different domains are injected into the testing phase, most open-set models' performances are not as robust as we assumed. Most of the results are task-dependent. Open-set methods such as OpenMax and OLTR use the source data to model the open-space risk, so there is a strong correlation between the open-space data distribution similarity to the source dataset and the expected effectiveness of the model to recognise open-set conditions. MC dropout and ODIN do not explicitly model the source data distribution for open space, which leads to a more consistent performance in rejecting unknowns.




\noindent\textbf{Uncertainty vs open-set focused models:}
There are two main types of open-set models we implemented in this study: uncertainty-based methods (MC dropout and ODIN) and open-set focused modelling (OpenMax and OLTR).
OpenMax and OLTR are open-set focused methods that have specific components to explicitly measure open-space risks.
To re-emphasise the point, uncertainty refers to the concept of noise inherent in the observations and models. We hypothesise that uncertainty estimation based on output variability brings the separation that a model is truly confident about, which may benefit the model in recognising out-of-distribution samples.
The epistemic uncertainty~\cite{begoli2019need} is addressed by the learning model, such as that from the MC dropout method implemented in this paper. In contrast to Bayesian MC dropout
approaches that perturb predictions to produce prediction distributions, ODIN simply uses temperature scaling in the softmax function and adds small controlled perturbations to the inputs.


From our observations of these two types of open-set models in three different scenarios, the uncertainty methods show competitive performance on recognising known conditions but obtain worse results than open-set focused models when detecting unknown conditions, especially on more challenging \textit{similar-domain} and \textit{different-domain} tasks.
Generally, we conclude that employing uncertainty training in the model and thresholding the neural networks' class confidence are normally not enough to reject unknowns.
These observations may be because unknowns do not necessarily appear as uncertain in the representation space, 
and uncertain inputs in the Bayesian approach do not prevent misclassification from occurring.


\noindent\textbf{Practical usage:}
In a real clinical workflow, real-time feedback is key for making machine learning models efficient. In Table~\ref{tab:speed}, we supply extra results on the various models' inference speed to make the diagnosis. The ODIN approach is employed as the baseline model with 1 unit of time (the actual speed depends on the network architecture and hardware environment, and we were able to run the feed-forward network in 112.3 ms with a GTX 1080Ti GPU, which has only 11 GB of memory). Class dictionaries such as the EVT methods OpenMax and OLTR are approximately 2 to 3 times slower but are still in an acceptable range. The Bayesian-based uncertainty method takes a much longer time for inference due to its distribution-dependent nature.
When taking model calibration performance into account (see Sec.~\ref{sec:calibra}), the OLTR approach is the most suitable deep CNN (DCNN)-based model for medical inference.

\begin{table}[t!]
	\centering
		\caption{Inference Efficiency}
\begin{tabular}{| c | c |}\hline
\backslashbox{Method}{Inference}
&\pbox{20cm}{Time Unit}\\[1ex]  \hline
\textbf{OpenMax} & $\times$2.8	 \\[1ex] \hline
\textbf{MC Dropout} & $\times$20	 \\[1ex] \hline
\textbf{ODIN} &  $\times$1 (112.3 ms)	 \\[1ex] \hline
\textbf{OLTR} & $\times$1.8	 \\[1ex] \hline

\end{tabular}
		\label{tab:speed}
\end{table}


\section{Conclusion}\label{sec:conclu}
In this study, we present rigorous evaluations of state-of-the-art open-set methods~\cite{bendale2016towards,liang2017enhancing,miller2018dropout,liu2019large}, explore three different open-set scenarios from the \textit{self-domain} to the \textit{different-domain} frameworks and compare them on various general and medical domain datasets, including dermatology, radiology and ophthalmology datasets. We have demonstrated how different scenarios will impact the open-set class recognition performance and confidence calibration of the various methods.
Our analysis and conclusion suggest that the target domain similarity from different scenarios should be quantised in the open space to assist in better representation learning. It is worthwhile to measure and use the estimated domain similarity~\cite{cui2018large} to guide model transfer learning in this task.
The source code and model logits needed to reproduce the results in the paper are available on the project's GitHub website. We hope it will provide common ground and good baselines for future benchmarking in this critical medical open-set field.

%
\IEEEpeerreviewmaketitle





\ifCLASSOPTIONcaptionsoff
  \newpage
\fi



%
\bibliographystyle{IEEEtran}
\bibliography{refs}

%








\end{document}